\documentclass[runningheads]{llncs}
\usepackage{graphicx}
\usepackage{xcolor}
\usepackage[utf8]{inputenc}
\usepackage{fancyvrb,fvextra}
\usepackage{enumitem}
\usepackage{xurl}
\usepackage[colorlinks, citecolor=blue]{hyperref}
\usepackage{caption}
\usepackage{tikz}
\usetikzlibrary{positioning, arrows.meta}
\usepackage{hyperref}

\tikzstyle{startstop} = [rectangle, rounded corners, minimum width=3cm, minimum height=1cm,text centered, draw=black, fill=red!30]
\tikzstyle{process} = [rectangle, minimum width=3cm, minimum height=1cm, text centered, draw=black, fill=blue!30]
\tikzstyle{arrow} = [thick,->,>=Stealth]

\begin{document}
\title{Benchmarking Different Application Types across Heterogeneous Cloud Compute Services}

\author{Nivedhitha Duggi, Masoud Rafiei, Mohsen Amini Salehi}
\institute{High Performance Cloud Computing (\href{http://hpcclab.org/}{HPCC}) Lab,\\ University of North Texas, Denton TX, USA\\ nivedhithaduggi@my.unt.edu and mohsen.aminisalehi@unt.edu}

\maketitle              
%
\begin{abstract}
Infrastructure as a Service (IaaS) clouds have become the predominant underlying infrastructure for the operation of modern and smart technology. IaaS clouds have proven to be useful for multiple reasons such as reduced costs, increased speed and efficiency, and better reliability and scalability. Compute services offered by such clouds are heterogeneous---they offer a set of architecturally diverse machines that fit efficiently executing different workloads. However, there has been little study to shed light on the performance of popular application types on these heterogeneous compute servers across different clouds. Such a study can help organizations to optimally (in terms of cost, latency, throughput, consumed energy, carbon footprint, etc.) employ cloud compute services. At HPCC lab, we have focused on such benchmarks in different research projects and, in this report, we curate those benchmarks in a single document to help other researchers in the community using them. Specifically, we introduce our benchmarks datasets for three application types in three different domains, namely: Deep Neural Networks (DNN) Inference for industrial applications, Machine Learning (ML) Inference for assistive technology applications, and video transcoding for multimedia use cases.
\end{abstract}

\section{Overview}
Heterogeneous computing systems (HCS) have become essential in overcoming computational limitations as Moore's Law reaches its twilight, especially with the rising demand for scalable and efficient computational solutions for smart applications. These systems leverage architecturally diverse machines, enabling them to handle tasks with varying computational requirements. Heterogeneity is increasingly prevalent across cloud systems, including public cloud platforms like Amazon Web Services (AWS) and Microsoft Azure, which provide a mix of x86-based CPUs, ARM processors, GPUs, and domain-specific accelerators like FPGAs.

On the application side, cloud computing has become a cornerstone for deploying a wide range of applications, including Deep Neural Network (DNN) inference and training tasks, and more conventional workloads, such as video processing, across various industries. These systems are highly diverse (aka application level heterogeneity) and have affinity with heterogeneous compute servers offered by cloud platforms. 

This technical report delves into the benchmarking of different application types across different cloud providers with heterogeneity compute offerings.
This benchmarking provides a base to evaluate and compare the performance of diverse workloads. In this paper, we focus on benchmarking applications for inference tasks across heterogeneous servers. Specifically, we analyze:

\begin{itemize}
\item Deep Neural Network (DNN) inference for time-critical applications in the Oil and Gas industry \cite{razin2020hpcc,razinfgcs}.
\item Machine Learning inference tasks, including image classification, NLP, and speech recognition for assistive technology, particularly, for blind and visually impaired users \cite{ucc24heet}.
\item Video transcoding tasks involving resolution, frame rate, bit rate, and codec conversions \cite{chavitTPDS21,shangrui20}.
\end{itemize}
This technical report introduces the methodologies, datasets, and cloud platforms used for benchmarking and provides insights into accessing these benchmarks across heterogeneous systems. The datasets of this report are all available on the High-Performance Cloud Computing Lab's GitHub repositories.

\subsection{Benchmarking Structure}
\begin{figure}
    \centering
    \includegraphics[width=1\linewidth]{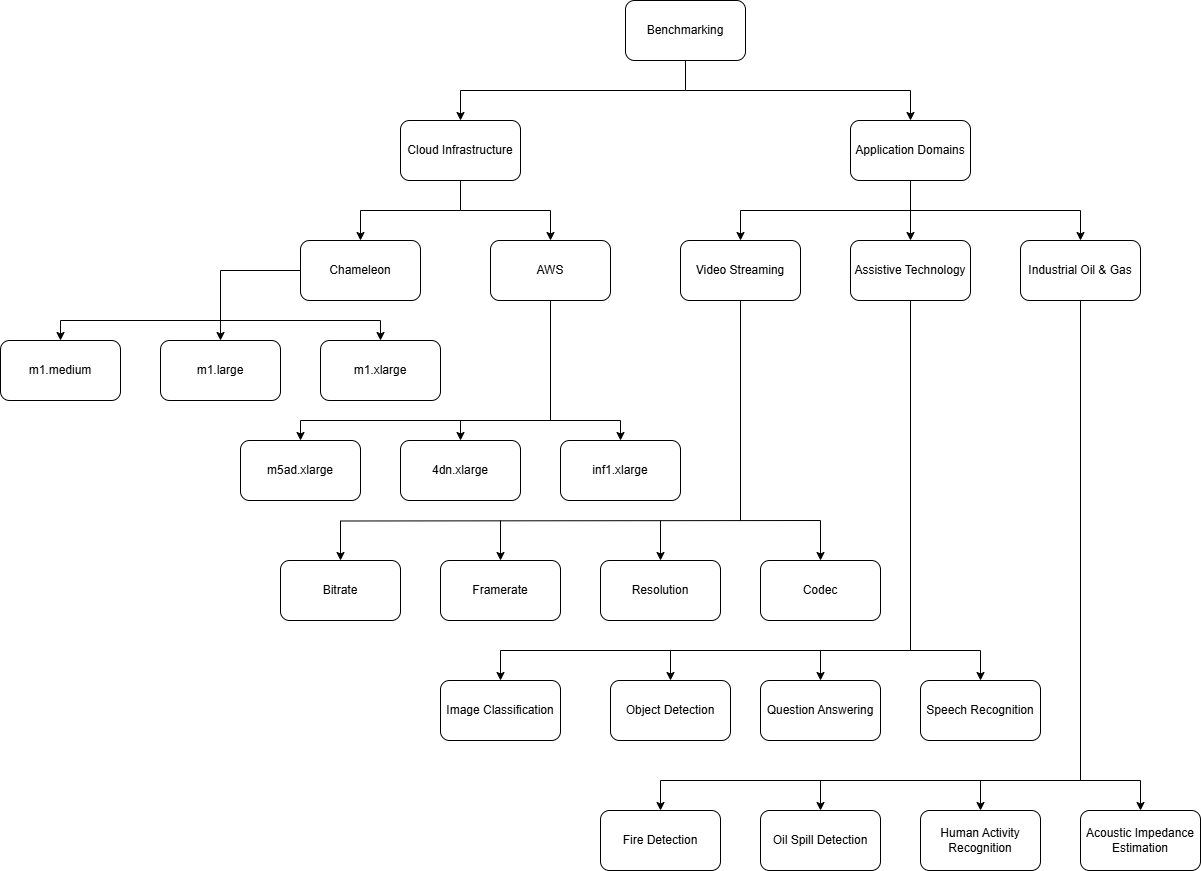}
    \caption{Benchmarking structure}
    \label{fig:1}
\end{figure}

Figure \ref{fig:1} provides an overview to the benchmarking we introduce in this technical report. According to the figure, our benchmarking encompasses two Cloud providers, namely AWS and Chameleon Cloud; each one with heterogeneous processors that are listed in the overview taxonomy. On the application side we consider three application domains, namely ``Video Transcoding'', ``Machine Learning'' inference for assistive technology, and ``Machine Learning'' inference for industrial Use Cases in the context of Oil and Gas. Each application is benchmarked on each processor type in different clouds.

\subsection{Application Types}
The industrial use cases include applications like fire detection, oil spill detection, and acoustic impedance estimation in the Oil and Gas sector, which require real-time and highly accurate predictions to prevent catastrophic failures.
   
Assistive Technology applications are in the context of supporting blind and visually impaired users, focusing on speech recognition and object (obstacle) detection to enhance accessibility and independence.

Video Processing application are for video transcoding and compression in multimedia streaming. The applications include changing resolution, bit rate, frame rate, and compression format.

\subsection{Heterogeneous Resources in Cloud Platforms}

Leading providers like Amazon Web Services (AWS) and research-oriented platforms such as Chameleon Cloud, offering a range of virtual machine (VM) configurations to cater to different application needs with different prices.

In this benchmarking we use VMs configured with different processor types offered by these cloud providers. 

Performance variability in heterogeneous systems poses significant challenges, especially for inference tasks where latency and throughput are critical. By analyzing execution times across varied machine types, we can figure out: (A) the randomness in application performance due to heterogeneity. (B) Comprehensive benchmarking results, including resource utilization and execution times. These results can be used by researchers to explore strategies to optimize resource allocation for cost-efficiency and scalability. (C) This benchmarking provides access to essential resources through curated links to GitHub repositories and datasets, enabling reproducibility and further research

In the rest of this technical report, we will explain details of benchmarking for industrial use case (Section~\ref{sec:og}), assistive technology for blind and visually-impaired users (Section~\ref{sec:bvi}), and video processing (Section~\ref{sec:video}).



\section{Benchmark I: DNN Applications in Industrial (Oil \& Gas) Use Case}\label{sec:og}

\subsection{Overview}
In the context of O\&G operations, several activities, such as fire detection, toxic gas monitoring, and spill detection, require accurate and timely processing. A failure to process these operations correctly and quickly can lead to disastrous consequences, including oil spills, explosions, and loss of life. Understanding the uncertainties in the execution times of different applications and properly modeling them is crucial for improving the safety, reliability, and efficiency of O\&G operations in Industry 4.0.

We benchmark the inference time of four Deep Neural Networks (DNN) applications in the context of O\&G industry on both AWS and Chameleon cloud, as explained in the next subsections. We note that more extensive results about this benchmark is available in \cite{razin2020hpcc}. This benchmark was also used in conducting the research and evaluation of \cite{razinfgcs}.

\subsection{DNN Inference Time for O\&G Applications}

\begin{enumerate}
\item \textbf{Fire Detection:} Utilizing the FireNet DNN model(which itself is based on Alexnet \cite{alom2018history}), this application detects fire in real-time from video frames. FireNet integrates convolutional layers with max-pooling and normalization to optimize processing time and accuracy. The dataset consists of 240 videos processed on heterogeneous cloud systems, including AWS and Chameleon Cloud.

All the benchmarking results and scripts for the Fire Detection Application are available in \href{https://github.com/hpcclab/Benchmarking-DNN-applications-industry4.0/tree/master/Applications/FireDetection}{GitHub Repository for Fire Detection}.

\item \textbf{Oil Spill Detection:} 
We utilize a detection system that operates based on the FCN-8 model \cite{long2015fully}. The model contains five Fully Convolutional Network (FCN) blocks and two up-sampling blocks that collectively perform semantic segmentation (i.e., classifying every pixel) of an input image and output a labeled image. The FCN-8 model functions based on the satellite (a.k.a. SAR) \cite{huang2020classification} images. We configure the analysis to obtain the inference time of 110 SAR images collected by MKLab \cite{krestenitis2019oil}.

All the benchmarking results and scripts for the Fire Detection Application are available in \href{https://github.com/hpcclab/Benchmarking-DNN-applications-industry4.0/tree/master/Applications/OilSpillDetection}{GitHub link for Oil Spill Detection}.

\item \textbf{Human Activity Recognition (HAR):} Human Activity Recognition (HAR) systems are widely used in Industry 4.0 to ensure workers safety in hazardous zones. For this purpose, motion sensors, such as accelerometer and gyroscope, that are widely available on handheld PDA devices are utilized. The HAR system we use operates based on the sequential neural network model with four layers to identify the worker's activities (namely, walking, walking upstairs, walking downstairs, sitting). For analysis, we use a dataset of UCI machine learning repository, known as Human Activity Recognition Using Smartphones \cite{anguita2013public}.

All the benchmarking results and scripts for this Application are available in \href{https://github.com/hpcclab/Benchmarking-DNN-applications-industry4.0/tree/master/Applications/HAR}{GitHub Repository for Human Activity Recognition}.

\item \textbf{Acoustic Impedance Estimation:} 
Estimating acoustic impedance (AI) from seismic data is an important step in O\&G exploration. To estimate AI from seismic data, we utilize a solution functions based on the temporal convolutional network \cite{mustafa2019estimation}. Marmousi 2 dataset \cite{marmousi} is used to estimate AI.

All the benchmarking results and scripts for this Application are available in \href{https://github.com/hpcclab/Benchmarking-DNN-applications-industry4.0/tree/master/Applications/AcqImp}{GitHub Repository for Acoustic Impedance Estimation}.

\end{enumerate}

\subsection{Cloud Platforms Used for Benchmarking the Industrial Applications}

From \textbf{Amazon Web Services (AWS)}, we employed the following instance type virtual machines (VMs):

\begin{enumerate}
\item \textbf{General Purpose Instances (e.g., m5ad.xlarge):} Balances compute, memory, and networking for diverse workloads.
\item \textbf{GPU Instances (e.g., g4dn.xlarge): }Optimized for high-performance tasks like deep learning and video processing.
\item \textbf{Inference-Optimized Instances (e.g., inf1.xlarge):} Designed for machine learning inference at scale. 

For configurations and raw data of the benchmarking under these AWS instances, please refer to \href{https://github.com/hpcclab/Benchmarking-DNN-applications-industry4.0/tree/master/AWS_Result}{AWS results for industrial use cases}.
\end{enumerate}

\textbf{Chameleon Cloud} is a configurable experimental platform that enables extensive benchmarking research. It offers VM flavors such as \texttt{m1.medium, m1.large}, and \texttt{m1.xlarge}, catering to various resource needs \cite{chameleon}. 

For configurations and raw data of the benchmarking under these Chameleon Cloud instances, please refer to \href{https://github.com/hpcclab/Benchmarking-DNN-applications-industry4.0/tree/master/Chameleon_Result}{Chameleon results for industrial use cases}.

\subsection{Benchmarking Methodology for Industrial Applications on Heterogeneous Clouds}

To benchmark the DNN applications, the inference time is measured in AWS and Chameleon cloud providers with respect to different machine types they offer. That is, the aforementioned applications are deployed across different cloud instances mentioned earlier. The random nature of the inference time is captured by running each application multiple times under varying cloud conditions to obtain a comprehensive understanding of how the system’s heterogeneity influences the performance.



Detailed benchmarking data are available in: \href{https://github.com/hpcclab/Benchmarking-DNN-applications-industry4.0}{Benchmarking Repository}.

   \subsection{Analysis of Inference Time of the Applications}
The inference time of each application will be analyzed using both AWS and Chameleon Cloud, and the times across the different machines will be compared. First, we will test for the normality of the datasets using the Shapiro-Wilk test, then we will utilize other statistical inference methods such as mean and standard deviations to compare the execution times on the different cloud services.

\subsubsection{Shapiro-Wilk Test for Normality of the Data}
The Shapiro-Wilk test is a test that evaluates whether the data set is normally distributed. A large p-value indicates that the dataset is normally distributed, while a small p-value indicates that the data is not normally distributed. The normality tests are used to understand the distribution of the collected data; therefore, this test is used.

Results  of the AWS Cloud Shapiro-Wilk Test; Raw data can be found at: \href{https://github.com/hpcclab/Benchmarking-DNN-applications-industry4.0/tree/master/AWS_Result}{Results  of the AWS Cloud Shapiro-Wilk Test}
        
Results  of the Chameleon Cloud Shapiro-Wilk Test; Raw data can be found at: \href{https://github.com/hpcclab/Benchmarking-DNN-applications-industry4.0/tree/master/Chameleon_Result}{Results  of the Chameleon Cloud Shapiro-Wilk Test}

\subsubsection{Kolmogorov-Smirnoff Goodness of Fit Test}
Due to the lack of normality in several cases after using the Shapiro-Wilk test, we utilized the Kolmogorov-Smirnoff test to determine the best fitting distribution of the inference times. The Kolmogorov-Smirnoff Goodness of Fit test identifies whether a set of samples derived from a population fits to a specific distribution.

Results  of the Chameleon Cloud Shapiro-Wilk Test; Raw data can be found at: \href{https://github.com/hpcclab/Benchmarking-DNN-applications-industry4.0/tree/master/Chameleon_Result}{Results  of the Chameleon Cloud Shapiro-Wilk Test}

\subsubsection{Mean and Standard Deviation of Inference Executive Times}
The mean and standard deviation of the inference times summarizes the behavior of the observations in a single value.
   
Results  of the AWS Mean and Standard Deviation; Raw data can be found at: \href{https://github.com/hpcclab/Benchmarking-DNN-applications-industry4.0/tree/master/AWS_Result}{Results  of the AWS Mean and Standard Deviation}

Results of the Chameleon Mean and Standard Deviation. Raw data can be found at: \href{https://github.com/hpcclab/Benchmarking-DNN-applications-industry4.0/tree/master/Chameleon_Result}{Results of the Chameleon Mean and Standard Deviation}

For further analysis of these results, please check our paper dedicated to this benchmarking \cite{razin2020hpcc}.

\section{Benchmark II: Machine Learning Inference Benchmarking on Heterogeneous Cloud Resources for Assistive Technology}\label{sec:bvi}
\subsection{Overview}
In the context of assistive technology, several activities, such as identifying surrounding object, obstacle detection, and interaction with the system (in form of speech recognition and question answering) require accurate and real-time processing. This benchmarking aims at Understanding the inference time of different applications for assistive technology of blind and visually impaired individuals.

We benchmark the inference time of four ML applications in the context of assistive technology on different VM types of AWS cloud, as explained in the next subsections. We note that this benchmark has been used and explained in our prior publication \cite{ucc24heet}.

 \subsection{Task Types}
In this section, we benchmark the execution times of four ML tasks used for assisting blind and visually impaired individuals, namely Image Classification, Object Detection, Question Answering, and Speech Recognition. To ensure consistency across the task types, they will all be converted to the ONNX (Open Neural Network Exchange) format using the Python ONNX converter. The workload of the tasks was also designed to be available for execution from the beginning. 

The summary of the benchmarking of these applications on different AWS VM types are shown in Table~\ref{tab:allML}.

        \begin{table}[]
        \centering
        \begin{tabular}{|c|c|c|c|}
            \hline
            Task Type & t2.large & c5.2xlarge & g4dn.xlarge \\
            \hline
            Image Classification \cite{he2016deep} & 74 & 27 & 12 \\
            Object Detection \cite{Jocher_YOLOv5_by_Ultralytics_2020} & 36 & 21 & 12 \\
            Question Answering \cite{sanh2019distilbert} & 36 & 16 & 4 \\
            Speech Recognition \cite{baevski2020wav2vec} & 621 & 237 & 20 \\
            \hline
        \end{tabular}
        \caption{Table showing the expected execution times for the different tasks across the different machines. The table is also available in the \href{https://github.com/hpcclab/heterogeneity_measure/blob/main/analysis/results/eets/eet-summarized.png} {ML Github repository}.}
        \label{tab:allML}
    \end{table}
    
    \subsubsection{Image Classification}
Image classification \cite{he2016deep} is a supervised learning problem where a model is trained to identify specific target classes using labeled example images. The aim is to classify unseen images accurately by learning patterns from the training data. In this dataset, image classification is implemented using the ResNet50 model, which is a robust deep neural network for image recognition tasks. To measure performance, 1,000 sample images are processed on each machine (VM) type to evaluate execution times for the classification task. 

\begin{figure}
        \centering
        \includegraphics[width=0.6\linewidth]{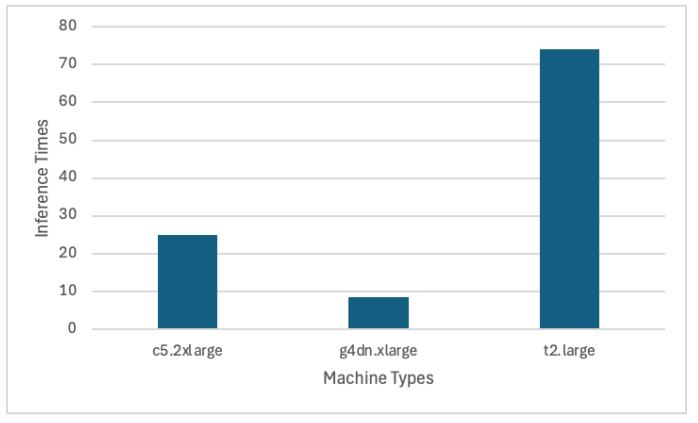}
        \caption{Chart comparing the Image Classification Inference times across the Machines}
        \label{fig:classification-Chart}
    \end{figure}

The average execution (inference) time of this model on different AWS VM types are shown in Table~\ref{tab:class}. Raw data of this benchmark can be found at \href{https://github.com/hpcclab/heterogeneity_measure/tree/main/analysis/results/profiling/image_classification}{Image Classification Github page}, and figure~\ref{fig:classification-Chart} graphically shows these different inference times.
    \begin{table}[]
        \centering
        \begin{tabular}{|c|c|}
        \hline
            Machine Type & Average Inference Time \\
            \hline
           c5.2xlarge & 24.929 \\
           g4dn.xlarge & 8.663 \\
           t2.large & 74.096 \\
        \hline
        \end{tabular}
        \caption{Average inference times of the three machine types when performing image classification tasks. }
        \label{tab:class}
    \end{table}
    
    \subsubsection{Object Detection}
Object detection \cite{Jocher_YOLOv5_by_Ultralytics_2020} is to locate and identify instances of objects within images or videos. In this dataset, YOLOv5 model is utilized to perform object detection, a task run on 1,000 sample images ten times to measure the expected execution time for object recognition. The set of images used for this purpose are available \href{https://drive.google.com/drive/folders/1lrlEzL2XFhmECxevFFXJgbP7lSQUy2fT?usp=sharing}{in this link}.

\begin{figure}
        \centering
        \includegraphics[width=0.6\linewidth]{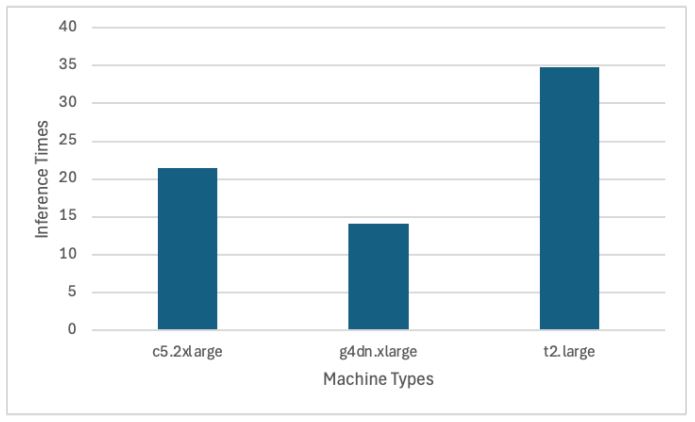}
        \caption{The average inference times for Object Detection across the different AWS VM types.}
        \label{fig:obj}
    \end{figure}

The average execution (inference) time of this model on different AWS VM types are shown in Table~\ref{tab:obj}. Raw data of this benchmark can be found at \href{https://github.com/hpcclab/heterogeneity_measure/tree/main/analysis/results/profiling/object_detection}{Object Detection Github page}, and figure~\ref{fig:obj} graphically shows these different inference times.
     \begin{table}[]
        \centering
        \begin{tabular}{|c|c|}
        \hline
           Machine & Average Inference Time \\
           \hline
           c2.2xlarge & 21.483 \\
           g4dn.xlarge & 14.149 \\
           t2.large & 38.824 \\
           \hline
        \end{tabular}
        \caption{The Average inference times of the three machines when performing the Object Detection. }
        \label{tab:obj}
    \end{table}
    
    \subsubsection{Question Answering}
    Question-Answering (QA)  models are advanced machine and deep learning models designed to answer questions based on provided context, or in some cases, without context, as seen in open-domain QA tasks. 
    
    In this dataset, the QA task is implemented using the DistilBERT model \cite{sanh2019distilbert}, a lightweight variant of BERT. A sample context and question are provided as input, and the inference task is executed 1,000 times. 

\begin{figure}
        \centering
        \includegraphics[width=0.6\linewidth]{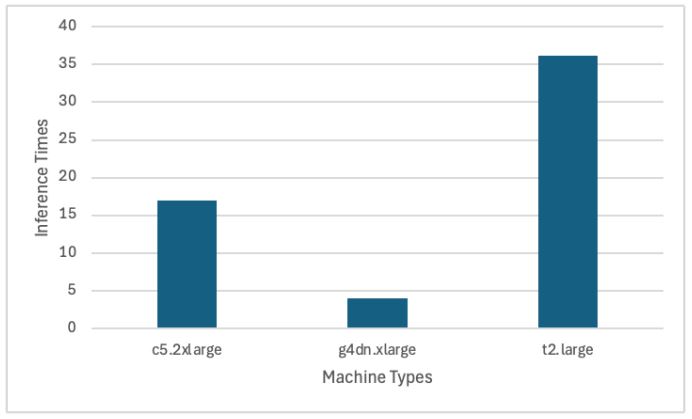}
        \caption{Average inference times of different machines performing questions answering tasks.}
        \label{fig:qa}
    \end{figure}

The average execution (inference) time of this model on different AWS VM types are shown in Table~\ref{tab:qa}. Raw data of this benchmark can be found at \href{https://github.com/hpcclab/heterogeneity_measure/tree/main/analysis/results/profiling/question_answering}{Question Answering Github page}, and figure~\ref{fig:qa} graphically shows these different inference times.
    
    \begin{table}[]
        \centering
        \begin{tabular}{|c|c|}
            \hline
            Machine Type & Average Inference Time \\
            \hline
            c2.2xlarge & 16.995 \\
            g4dn.xlarge & 4.035 \\
            t2.large & 36.116 \\
            \hline
        \end{tabular}
        \caption{Average inference time of performing the question-answering inference across heterogeneous AWS VM types.}
        \label{tab:qa}
    \end{table}
    
    \subsubsection{Speech Recognition}
    Automatic speech recognition (ASR) is the technology that enables programs to convert spoken language into written text. It differs from voice recognition, which focuses solely on identifying individual speakers. 

    In this dataset, speech recognition tasks are performed using the Wav2vec2 model \cite{baevski2020wav2vec}. Four-second audio samples are executed across various machine types, and the average inference time is calculated to evaluate the system's performance.

\begin{figure}
        \centering
        \includegraphics[width=0.6\linewidth]{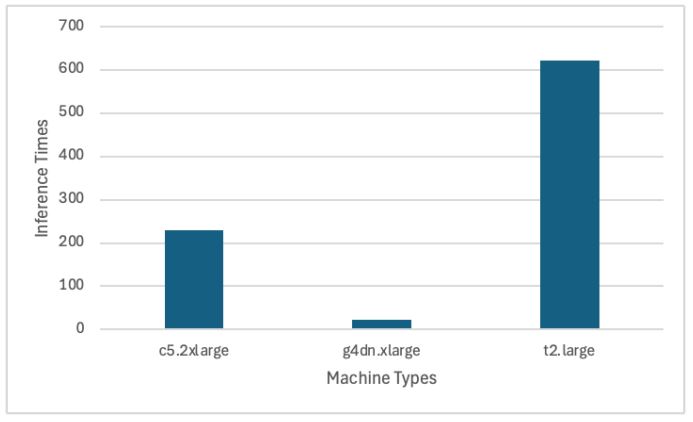}
        \caption{Average inference times of heterogeneous AWS machines performing speech recognition.}
        \label{fig:asr}
    \end{figure}

    The average execution (inference) time of this model on different AWS VM types are shown in Table~\ref{tab:asr}. Raw data of this benchmark can be found at \href{https://github.com/hpcclab/heterogeneity_measure/tree/main/analysis/results/profiling/speech_recognition}{Speech Recognition Github page}, and figure~\ref{fig:asr} graphically shows these different inference times.

    \begin{table}[]
        \centering
        \begin{tabular}{|c|c|}
            \hline
            Machine Types & Average Inference Times \\
            \hline
            c2.2xlarge & 230.32 \\
            g4dn.xlarge & 21.579 \\
            t2.large & 621.614 \\
            \hline
        \end{tabular}
        \caption{Average inference times of heterogeneous AWS VM types performing the speech recognition task.}
        \label{tab:asr}
    \end{table}
    
\section{Benchmark III: Video Transcoding Benchmark on Heterogeneous Cloud Resources}\label{sec:video} \subsection{Overview}
In this section, we explain the benchmark we produced for video processing. 
The benchmarking focuses on the video transcoding operation, which CPU-intensive, using FFMpeg, a popular tool for video processing. We note that FFMpeg is a powerful tool with the ability to concurrently process multiple operations. Therefore, the benchmarking includes multi-parameter processing information on the cloud, in addition to single parameter transcoding. 

This benchmark has been used in \cite{shangrui20} and \cite{chavitTPDS21} publications. Interested researchers can refer to these papers for more details on the specifications of the benchmark and how to use it. Here is the \href{https://github.com/hpcclab/videostreamingBenchmark?tab=readme-ov-file}{Github repository} where all the benchmarking results and scripts are made available. This Github repositoryincludes a \href{https://github.com/hpcclab/videostreamingBenchmark/tree/master/VideoPreparationScripts}{folder} where all the scripts for splitting and transcoding benchmark videos are included. 

\subsection{Video Transcoding Tasks}
This repository includes 100 open-source videos that are gathered in a variety of content forms (i.e. fast/slow pace, action, scenery, animation, etc.). All 100 videos are available in the following formats 
\begin{enumerate}
    \item \href{https://drive.google.com/drive/folders/1uereCYUqTqb602W9BFi-cjj-Gag-IFt9?usp=sharing}{In their original form}. Also, in this \href{https://drive.google.com/file/d/1Z9g91wXzOIlFM7pWEaLkDq0Wc7-SGAaQ/view?usp=sharing}{link} the meta-data (content type, codec, resolution,  of each video is described in the CSV format.
    \item \href{https://drive.google.com/drive/folders/1MaEAN8TjuOhv9mH33j5L7nibxppriadQ?usp=sharing}{In form of 2-second split video segments}.
    \item The 2-second split video segments in \href{https://drive.google.com/drive/folders/1KhsxZtC22L-EHoeXmsdmWpkuNZpmS-pL?usp=sharing}{the uniform (standard) compression format}. We define standard as to have a 720p resolution, 30 fps framerate, H.264 codec, and 4500 kb/s bitrate
    \item the 2-second split video segments available with the \href{https://drive.google.com/drive/folders/1MKeNOcfzrWl9kUNp26F-5eFm4qwhpjsE?usp=sharing}{standard and HLS-compatible format}. 
\end{enumerate}

The task types we considered for this benchmarking ar the following three transcoding operations:
    
\subsubsection{Bitrate}
    Bitrate describes the rate at which bits are transferred from one location to another along a digital network. The average execution time is collected for five parameters: 384K, 512K, 768K, 1024K, and 1536K. 
    
    
    
    \subsubsection{Frame rate}
    Frame rate is the measurement of how quickly a number of frames appear in a second. The average execution time is collected for the five parameters: 10fps, 15fps, 20fps, 30fps, and 40fps. 
    
    
    \subsubsection{Resolution}
    Resolution is the total number of pixels in a video frame rate. The average execution time for the resolution tasks will be collected for the five parameters: 352$\times$288, 680$\times$320, 720$\times$480, 1280$\times$800, and 1920$\times$1080. 
    
\subsection{Single Parameter Video Transcoding Benchmarking}


To limit the degree of freedom in execution time, in this part, each task is configured to change only one specification of the videos in the benchmark dataset.
\href{https://github.com/hpcclab/videostreamingBenchmark/tree/master/ExecutionTimeBenchmark/SingleParameterTranscoding}{Single Parameter Transcoding folder} include transcoding which changes one parameter of the standardized video segment at a time; the folder also includes the required scripts to perform such benchmarking. All the results of this benchmark are from processing the video dataset only on \textbf{Chameleon Cloud Small VM} type. Each experiment is run 30-times. Both individual transcoding time and the mean and standard deviation are reported in the benchmark.

\subsection{Multi-Parameter Video Transcoding Benchmarking}
\begin{figure}
        \centering
        \includegraphics[width=1\linewidth]{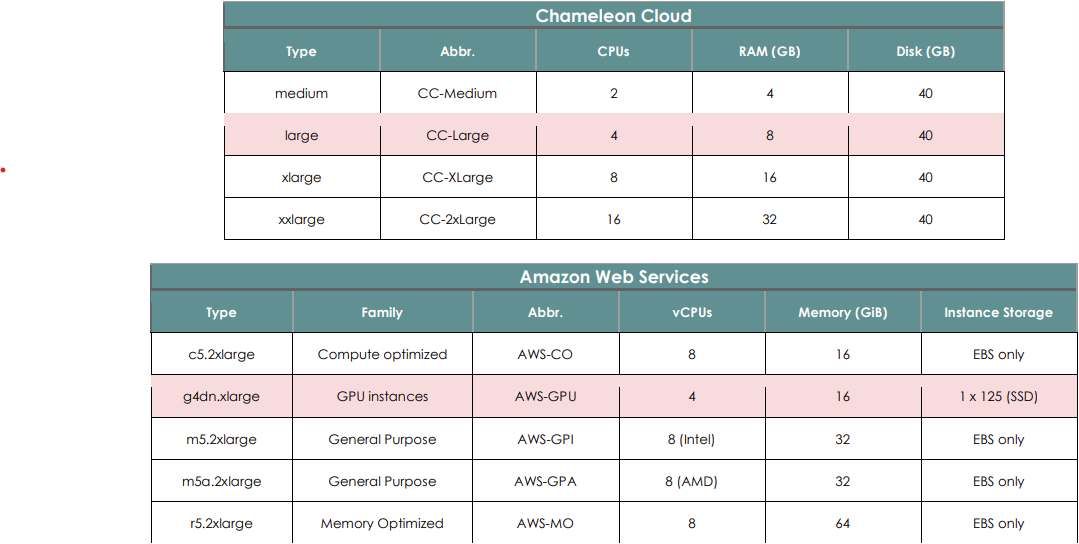}
        \caption{Different machine types used for benchmarking multi-parameter video transcoding. Image taken for the main report file available in \href{https://github.com/hpcclab/videostreamingBenchmark/blob/master/ExecutionTimeBenchmark/MultipleParameterTranscoding/SamplingAnalysis/Nguyen_Diana_CMPS490_Report.pdf}{here}.}
        \label{fig:cloudres}
    \end{figure}

\href{https://github.com/hpcclab/videostreamingBenchmark/tree/master/ExecutionTimeBenchmark/MultipleParameterTranscoding}{Multiple Parameter Transcoding folder} include transcoding which using FFMpeg, we change three  parameters, namely Framerate, Resolution, Codec (e.g., \texttt{30, 240p, HEVC}) of the standardized video segment in one transcoding operation (i.e., output one video with three parameters changed). This benchmarking was conducted both on the following VM types of Chameleon Cloud: \texttt{Medium, Large, XLarge, XXLarge};
For AWS cloud we used \texttt{c5.2xlarge, g4dn.xlarge, m5.2xlarge, m5a.2xlarge, r5.2xlarge} VM types. More information about these resources are shown in the tables of Figure \ref{fig:cloudres} These results are available in separate folders. 

\begin{figure}
        \centering
        \includegraphics[width=1\linewidth]{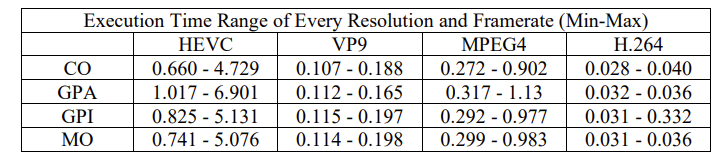}
        \caption{Min and Max of transcoding time for different machine types ins AWS. CO is \texttt{AWS c5.2xlarge}, GPA is \texttt{AWS m5a.2xlarge} with AMD processor, GPI is \texttt{AWS m5.2xlarge} with Intel processor, and MO is \texttt{AWS r5.2xlarge} . Image taken for the main report file available in \href{https://github.com/hpcclab/videostreamingBenchmark/blob/master/ExecutionTimeBenchmark/MultipleParameterTranscoding/SamplingAnalysis/Nguyen_Diana_CMPS490_Report.pdf}{Page 45}.}
        \label{fig:awsres}
    \end{figure}

The summary of analysis of multi-parameter transcoding for different codec types for AWS cloud is show in the table of Figure~\ref{fig:awsres}. Each entry of the table shows the minimum and maximum transcoding time of the resolution, frame rate, and/or bit rate change for different  compression formats (codec). 

\subsection{Merging Multiple Transcoding Benchmarking}
FFMpeg has the ability to produce multiple outputs in one command (a.k.a. transcoding merging). Note that this is different with multi-parameter transcoding, because in merging, there are more than one output files. For instance, if we transcode frame rate, bit rate, and resolution of \texttt{video1}, and generate one output video file, this is a multi-parameter transcoding, whereas, changing the resolution of \texttt{video1} to \texttt{720p} and \texttt{1280p} and generate two separate video files is called merging (i.e., merging to different FFMpeg commands in one).

Early evaluation of the collected execution-time revealed a remarkable variation in the execution-time of some task types. Specifically, we noticed that codec execution-time is far beyond the other three task types.
Accordingly, we categorize the tasks types into two groups: \emph{First} group is called Video Information Conversion (\emph{VIC}) that includes changing bit-rate, frame-rate, or resolution task types. Tasks of this group have a low variation in their execution-times, when processing different video segments on the same machine type. \emph{Second} group is Video Compression Conversion that only includes the codec task type (hence, we call it the Codec group). In contrast to the first group, the codec execution-time (and subsequently its merge-saving) for different video segments varies remarkably even on the same machine.

This is the \href{https://github.com/hpcclab/videostreamingBenchmark/tree/master/MergingDataset}{link} to the merging benchmark and the paper elaborates on it is available here \cite{shangrui20}. 
The above Github page contains three sub folder, namely scripts, Codec class, and  (a.k.a. VIC) class. This dataset is prepared to show merge saving percentage (i.e., relative percentage between merge and non-merge counterpart).

The \textbf{Codec class} contains 3 subfolders which each of them have two Microsoft Excel files (xls format) that include the execution time of merged and non merged transcoding. Task merging in this case are performed with the same operation only. There is also ``merge across operations.xls'' that includes the data from where tasks from multiple operations within VIC class are merged together. The \textbf{VIC class} contains 3 subfolders where each one of them represents transcoding time of each of the codec (HEVC, mpeg, VP9). Both as a single task and a merged task.

The data collected in this benchmark are performed on our own local cloud which is a DELL PowerEdge R830 with 4×Intel Xeon E5-4628Lv4 processors with 112 homogeneous cores, 384 GB memory (24×16 GB DRAM), and RAID1 (2×900 GB HDD) storage.



\section{Summary}
This technical report serves as a handbook for different benchmarking we have done on various application types on different clouds with heterogeneous machines. We presented three main benchmarked applications in different domains:

\begin{enumerate}
    \item Four DNN applications in industrial use case (Oil and Gas context). This benchmarking was conducted on Chameleon and AWS clouds, each one with heterogeneous VMs.
    \item Four ML applications in assistive technology (blind and visually impaired) use case. This benchmarking was exclusively done on heterogeneous AWS VMs. 
    \item Four video transcoding operations using FFMpeg. This benchmarking was performed in three phases. First, single parameter transcoding on Chameleon Cloud. Second, Multi-parameter transcoding on both Chameleon and AWS clouds. Third, merging multiple transcoding operations in one FFMpeg command that was done on our local cluster. 
\end{enumerate}

All these benchmarks are publicly available through our Github and are reflected in our publications. We hope that other researchers in the community can make use of this benchmarks as well.




\bibliographystyle{ieeetr}
\bibliography{references}

\end{document}